
%
%

%
\def\reff{\par\noindent\hangindent 20pt}
%
\def\smallcap{\relax}
\def\sc{\relax}
\font\sevbf=cmbx7
\font\eightrm=cmr8
\font\eightbf=cmbx8
\font\ninerm=cmr9
\font\ninei=cmmi9
\font\nineit=cmti9
\font\ninesl=cmsl9
\font\ninebf=cmbx9
\font\ninesy=cmsy9
\def\rmnine{\fam0\ninerm}
\def\itnine{\fam\itfam\nineit}
\def\slnine{\fam\slfam\ninesl}
\def\bfnine{\fam\bffam\ninebf}
\def\ninepoint{\let\rm=\rmnine
\textfont0=\ninerm\scriptfont0=\sevenrm
\textfont1=\ninei\scriptfont1=\seveni
\textfont2=\ninesy
\textfont\itfam=\nineit \let\it=\itnine
\textfont\slfam=\ninesl \let\sl=\slnine
\textfont\bffam=\ninebf \scriptfont\bffam=\sevbf
\let\bf=\bfnine
\let\smallcap=\sevenrm
\let\sc=\sevenrm
\normalbaselineskip=10pt\normalbaselines\rm}
\font\tenib=cmmib10
\font\tenslbf=cmbxsl10

\def\rmten{\fam0\tenrm}
\def\itten{\fam\itfam\tenit}
\def\slten{\fam\slfam\tensl}
\def\bften{\fam\bffam\tenbf}
\def\tenpoint{\let\rm=\rmten
\textfont0=\tenrm\scriptfont0=\sevenrm
\textfont1=\teni\scriptfont1=\seveni
\textfont2=\tensy
\textfont\itfam=\tenit \let\it=\itten
\textfont\slfam=\tensl \let\sl=\slten
\textfont\bffam=\tenbf \scriptfont\bffam=\ninebf
\let\bf=\bften
\let\smallcap=\eightrm
\let\sc=\eightrm
\normalbaselineskip=12pt\normalbaselines\rm}
\def\rmbften{\fam0\tenbf}
\def\itbften{\fam\itfam\tenib}
\def\slbften{\fam\slfam\tenslbf}
\def\bftenpoint{\let\rm=\rmbften
\textfont0=\tenbf\scriptfont0=\sevbf
\textfont1=\tenib\scriptfont1=\seveni
\textfont2=\tensy
\textfont\itfam=\tenib \let\it=\itbften
\textfont\slfam=\tenslbf \let\sl=\slbften
\textfont\bffam=\tenbf \scriptfont\bffam=\ninebf
\let\bf=\bften
\let\smallcap=\eightbf
\let\sc=\eightbf
\normalbaselineskip=12pt\normalbaselines\rm}
\font\twelbf=cmb10 scaled \magstep1
\font\twelmib=cmmib10 scaled \magstep1
\font\twelslbf=cmbxsl10 scaled \magstep1
\font\fortbf=cmb10   scaled \magstep2
\font\fortsy=cmsy10   scaled \magstep2
\def\rmbffort{\fam0\fortbf}
\def\itbffort{\fam\itfam\twelmib}
\def\slbffort{\fam\slfam\twelslbf}
\def\bffort{\fam\bffam\fortbf}
\def\bffortpoint{\let\rm=\rmbffort
\textfont0=\fortbf  \scriptfont0=\twelbf
\textfont1=\twelmib \scriptfont1=\twelmib
\textfont2=\fortsy
\textfont\itfam=\twelmib  \let\it=\itbffort
\textfont\slfam=\twelslbf \let\sl=\slbffort
\textfont\bffam=\fortbf \scriptfont\bffam=\twelbf
\let\bf=\bffort
\let\smallcap=\twelbf
\let\sc=\twelbf
\normalbaselineskip=16pt\normalbaselines\rm}
%

\vsize=23.5truecm
\hoffset=-1true cm
\voffset=-1true cm
\newdimen\fullhsize
\fullhsize=40cc
\hsize=19.5cc
\def\fullline{\hbox to\fullhsize}
\def\makefootline{\baselineskip=10dd \fullline{\the\footline}}
\def\makeheadline{\vbox to 0pt{\vskip-22.5pt
            \fullline{\vbox to 8.5pt{}\the\headline}\vss}\nointerlineskip}
\let\lr=L \newbox\leftcolumn
\output={\global\topskip=10pt
         \if L\lr
            \global\setbox\leftcolumn=\columnbox \global\let\lr=R
            \message{[left\the\pageno]}%
            \ifnum\pageno=1
               \global\topskip=\fullhead\fi
         \else
            \doubleformat \global\let\lr=L
         \fi
         \ifnum\outputpenalty>-2000 \else\dosupereject\fi}
\def\doubleformat{\shipout\vbox{\makeheadline
    \fullline{\box\leftcolumn\hfil\columnbox}
           \makefootline}
           \advancepageno}
\def\columnbox{\leftline{\pagebody}}
\outer\def\bye{\sterne=1\ifx\speciali\undefined\else
\loop\smallskip\noindent special character No\number\sterne:
\csname special\romannumeral\sterne\endcsname
\advance\sterne by 1\global\sterne=\sterne
\ifnum\sterne<11\repeat\fi
\if R\lr\null\fi\vfill\supereject\end}
\hfuzz=2pt
\vfuzz=2pt
\tolerance=1000
\fontdimen3\tenrm=1.5\fontdimen3\tenrm
\fontdimen7\tenrm=1.5\fontdimen7\tenrm
\abovedisplayskip=3 mm plus6pt minus 4pt
\belowdisplayskip=3 mm plus6pt minus 4pt
\abovedisplayshortskip=0mm plus6pt
\belowdisplayshortskip=2 mm plus4pt minus 4pt
\predisplaypenalty=0
\clubpenalty=20000
\widowpenalty=20000
\parindent=1.5em
\frenchspacing
\def\newline{\hfill\break}%

\newtoks\vol
\newtoks\startpage \newtoks\lastpage
\newtoks\month \newtoks\year
\vol={1} %
\month={JANVIER}\year={2001}%
\startpage={1}\lastpage={99}%
\nopagenumbers
\def\paglay{\headline={
{\ninepoint\hsize\fullhsize\ifnum\pageno=1
\vtop{\baselineskip=10dd\hrule width52.5mm height0.4pt \kern5pt
\hbox{\noindent ASTRONOMY\ \ \&\ \ ASTROPHYSICS}\kern5pt
\hbox{SUPPLEMENT\ \ SERIES}
\kern5pt\hrule width52.5mm height0.4pt \kern5pt
\hbox{{\sl Astron. Astrophys. Suppl. Ser.}
{\bf \the\vol ,}\ \ \the\startpage -\the\lastpage\ \ (\the\year)}}
\hfill\vtop{\kern5pt\hbox{\the\month\ {\bf \the\year ,} PAGE \the\startpage}}
\else\ifodd\pageno --- \hfil--- --- ---\hfil\folio
\else \folio\hfil --- --- --- \hfil---\fi\fi}}}
\ifx \undefined\instruct
\headline={\tenrm\ifodd\pageno\hfil\folio
\else\folio\hfil\fi}\fi
\newcount\eqnum\eqnum=0
\def\autnum{\global\advance\eqnum by 1{\rm(\the\eqnum)}}

\def\utw{\smash{\rlap{\lower5pt\hbox{$\sim$}}}}
\def\udtw{\smash{\rlap{\lower6pt\hbox{$\approx$}}}}

\def\diameter{{\ifmmode\mathchoice
{\ooalign{\hfil\hbox{$\displaystyle/$}\hfil\crcr
{\hbox{$\displaystyle\mathchar"20D$}}}}
{\ooalign{\hfil\hbox{$\textstyle/$}\hfil\crcr
{\hbox{$\textstyle\mathchar"20D$}}}}
{\ooalign{\hfil\hbox{$\scriptstyle/$}\hfil\crcr
{\hbox{$\scriptstyle\mathchar"20D$}}}}
{\ooalign{\hfil\hbox{$\scriptscriptstyle/$}\hfil\crcr
{\hbox{$\scriptscriptstyle\mathchar"20D$}}}}
\else{\ooalign{\hfil/\hfil\crcr\mathhexbox20D}}%
\fi}}
\normalbaselines\rm
\def\petit{\def\rm{\fam0\eightrm}
\textfont0=\eightrm \scriptfont0=\sixrm \scriptscriptfont0=\fiverm
 \textfont1=\eighti \scriptfont1=\sixi \scriptscriptfont1=\fivei
 \textfont2=\eightsy \scriptfont2=\sixsy \scriptscriptfont2=\fivesy
 \def\it{\fam\itfam\eightit}%
 \textfont\itfam=\eightit
 \def\sl{\fam\slfam\eightsl}%
 \textfont\slfam=\eightsl
 \def\bf{\fam\bffam\eightbf}%
 \textfont\bffam=\eightbf \scriptfont\bffam=\sixbf
 \scriptscriptfont\bffam=\fivebf
 \def\tt{\fam\ttfam\eighttt}%
 \textfont\ttfam=\eighttt
 \let\tams=\kleinhalbcurs
 \let\tenbf=\eightbf
 \let\sevenbf=\sixbf
 \normalbaselineskip=9dd
 \if Y\REFERE \normalbaselineskip=2\normalbaselineskip
 \normallineskip=2\normallineskip\fi
 \setbox\strutbox=\hbox{\vrule height7pt depth2pt width0pt}%
 \normalbaselines\rm}%
\def\begpet{\vskip6pt\bgroup\ninepoint}
\def\endpet{\vskip6pt\egroup}
\newbox\FigT
\def\rahmen#1{
\setbox\FigT=\vbox{\hbox to4true cm{
\hfil\vrule width 0.01pt height 2cm\hfil}
\hrule height 0.01pt}
\vbox to#1true cm{\vfil\line{\hfill\box\FigT\hfill}}}
\def\begfig#1cm#2\endfig{\par
   \ifvoid\topins\midinsert\bigskip\vbox{\rahmen{#1}#2}\endinsert
   \else\topinsert\vbox{\rahmen{#1}#2}\endinsert
\fi}
\def\begfigwid#1cm#2\endfig{\par
\if N\lr\else
\if R\lr
\shipout\vbox{\makeheadline
\line{\box\leftcolumn}\makefootline}\advancepageno
\fi\let\lr=N
\topskip=10pt
\output={\plainoutput}%
\fi
\topinsert\line{\vbox{\hsize=\fullhsize\rahmen{#1}#2}\hss}\endinsert}
\def\figure#1#2{\bigskip\noindent{\ninepoint F{\sc IGURE} #1.\
\ignorespaces #2\smallskip}}
\def\begtab#1cm#2\endtab{\par
   \ifvoid\topins\midinsert\medskip\vbox{#2\rahmen{#1}}\endinsert
   \else\topinsert\vbox{#2\rahmen{#1}}\endinsert
\fi}
\def\begtabemptywid#1cm#2\endtab{\par
\if N\lr\else
\if R\lr
\shipout\vbox{\makeheadline
\line{\box\leftcolumn}\makefootline}\advancepageno
\fi\let\lr=N
\topskip=10pt
\output={\plainoutput}%
\fi
\topinsert\line{\vbox{\hsize=\fullhsize#2\rahmen{#1}}\hss}\endinsert}
\def\begtabfullwid#1\endtab{\par
\if N\lr\else
\if R\lr
\shipout\vbox{\makeheadline
\line{\box\leftcolumn}\makefootline}\advancepageno
\fi\let\lr=N
\output={\plainoutput}%
\fi
\topinsert\line{\vbox{\hsize=\fullhsize\noindent#1}\hss}\endinsert}
\def\tabcap#1#2{\smallskip\noindent T{\sc ABLE} #1.\
\ignorespaces{\it #2}\bigskip}
\def\begfullpage{\vfill\supereject
            \if R\lr\null\vfill\supereject\fi
            \begingroup\output={\plainoutput}
            \hsize=\fullhsize}
\def\endfullpage{\vfill\supereject\endgroup\let\lr=L}
\def\begref{\begingroup\let\INS=N\let\refer=\ref}
\def\ref{\goodbreak\if N\INS\let\INS=Y\vbox{\bigskip\noindent\tenbf
References\bigskip}\fi\hangindent\parindent
\hangafter=1\noindent\ignorespaces}
\def\endref{\endgroup}


%
 \def \aTa  { \goodbreak
     \bgroup
     \par
     \rightskip=0pt plus2cm\spaceskip=.3333em \xspaceskip=.5em
     \pretolerance=10000
     \noindent
     \bffortpoint}
 %
 \def \eTa{\vskip10pt\egroup
     \noindent
     \ignorespaces}
%
 \def \aTb{\goodbreak
     \bgroup
     \par
     \rightskip=0pt plus2cm\spaceskip=.3333em \xspaceskip=.5em
     \pretolerance=10000
     \noindent
     \bftenpoint}
 %
 \def \eTb{\vskip10pt
     \egroup
     \noindent
     \ignorespaces}
%
\catcode`\@=11
\expandafter \newcount \csname c@Tl\endcsname
    \csname c@Tl\endcsname=0
\expandafter \newcount \csname c@Tm\endcsname
    \csname c@Tm\endcsname=0
\expandafter \newcount \csname c@Tn\endcsname
    \csname c@Tn\endcsname=0
\expandafter \newcount \csname c@To\endcsname
    \csname c@To\endcsname=0
\expandafter \newcount \csname c@Tp\endcsname
    \csname c@Tp\endcsname=0
\def \resetcount#1    {\global
    \csname c@#1\endcsname=0}
\def\@nameuse#1{\csname #1\endcsname}
\def\arabic#1{\@arabic{\@nameuse{c@#1}}}
\def\@arabic#1{\ifnum #1>0 \number #1\fi}
 %
\expandafter \newcount \csname c@fn\endcsname
    \csname c@fn\endcsname=0
\def \stepc#1    {\global
    \expandafter
    \advance
    \csname c@#1\endcsname by 1}
\catcode`\@=12
%
%
   \catcode`\@= 11
%
\def\footnoterule{\kern-3pt\hrule width 2true cm\kern2.6pt}
\newinsert\footins
\def\footnotea#1{\let\@sf\empty 
  \ifhmode\edef\@sf{\spacefactor\the\spacefactor}\/\fi
  {#1}\@sf\vfootnote{#1}}
\def\vfootnote#1{\insert\footins\bgroup
  \ninepoint
  \interlinepenalty\interfootnotelinepenalty
  \splittopskip\ht\strutbox 
  \splitmaxdepth\dp\strutbox \floatingpenalty\@MM
  \leftskip\z@skip \rightskip\z@skip \spaceskip\z@skip \xspaceskip\z@skip
  \textindent{#1}\footstrut\futurelet\next\fo@t}
\def\fo@t{\ifcat\bgroup\noexpand\next \let\next\f@@t
  \else\let\next\f@t\fi \next}
\def\f@@t{\bgroup\aftergroup\@foot\let\next}
\def\f@t#1{#1\@foot}
\def\@foot{\strut\egroup}
\def\footstrut{\vbox to\splittopskip{}}
\skip\footins=\bigskipamount 
\count\footins=1000 
\dimen\footins=8in 
   \def \bfootax  {\bgroup\tenrm
                  \baselineskip=12pt\lineskiplimit=-6pt
                  \hsize=19.5cc
                  \def\textindent##1{\hang\noindent\hbox
                  to\parindent{##1\hss}\ignorespaces}%
                  \footnotea{$^\star$}\bgroup}
   \def \efootax  {\egroup\egroup}
   \def \bfootay  {\bgroup\tenrm
                  \baselineskip=12pt\lineskiplimit=-6pt
                  \hsize=19.5cc
                  \def\textindent##1{\hang\noindent\hbox
                  to\parindent{##1\hss}\ignorespaces}%
                  \footnotea{$^{\star\star}$}\bgroup}
   \def \efootay  {\egroup\egroup }
   \def \bfootaz {\bgroup\tenrm
                  \baselineskip=12pt\lineskiplimit=-6pt
                  \hsize=19.5cc
                  \def\textindent##1{\hang\noindent\hbox
                  to\parindent{##1\hss}\ignorespaces}%
                 \footnotea{$^{\star\star\star}$}\bgroup}
   \def \efootaz {\egroup \egroup}
\def\fonote#1{\mehrsterne$^{\the\sterne}$\begingroup
       \def\textindent##1{\hang\noindent\hbox
       to\parindent{##1\hss}\ignorespaces}%
\vfootnote{$^{\the\sterne}$}{#1}\endgroup}
\catcode`\@=12
\everypar={\let\lasttitle=N\everypar={\parindent=1.5em}}%
%
%
\def \titlea#1{\stepc{Tl}
     \resetcount{Tm}
     \if A\lasttitle\else\vskip22pt\fi
     \ifdim\pagetotal>\pagegoal\else
     \setbox0=\vbox{
     \noindent\bf
     \rightskip 0pt plus4em
     \pretolerance=20000
     \arabic{Tl}.\
     \ignorespaces#1
     \vskip11pt}
     \dimen0=\ht0\advance\dimen0 by\dp0\advance\dimen0 by 4\baselineskip
     \advance\dimen0 by\pagetotal
     \ifdim\dimen0>\pagegoal\eject\fi\fi
     \bgroup
     \noindent
     \bf
     \rightskip 0pt plus4em
     \pretolerance=20000
     \arabic{Tl}.\
     \ignorespaces#1
     \vskip11pt
     \egroup
     \nobreak
     \parindent=0pt
     \everypar={\parindent=1.5em
     \let\lasttitle=N\everypar={\let\lasttitle=N}}%
     \let\lasttitle=A%
     \ignorespaces}
 \def\titleb#1{\stepc{Tm}
     \resetcount{Tn}
     \if N\lasttitle\else\vskip-11pt\vskip-\baselineskip \fi
     \vskip17pt
     \ifdim\pagetotal>\pagegoal\else
     \setbox0=\vbox{
     \raggedright
     \pretolerance=10000
     \noindent
     {\bf\arabic{Tl}}.\arabic{Tm}.\
     \ignorespaces#1
     \vskip8pt}
     \dimen0=\ht0\advance\dimen0 by\dp0\advance\dimen0 by 4\baselineskip
     \advance\dimen0 by\pagetotal
     \ifdim\dimen0>\pagegoal\eject\fi\fi
     \bgroup
     \raggedright
     \noindent \pretolerance=10000
     {\bf\arabic{Tl}}.\arabic{Tm}.\
     \ignorespaces#1
     \vskip8pt
     \egroup
     \nobreak
     \let\lasttitle=B%
     \parindent=0pt
     \everypar={\parindent=1.5em
     \let\lasttitle=N\everypar={\let\lasttitle=N}}%
     \ignorespaces}
 \def \titlec#1{\stepc{Tn}
     \resetcount{To}
     \if N\lasttitle\else\vskip-3pt\vskip-\baselineskip\fi
     \vskip11pt
     \ifdim\pagetotal>\pagegoal\else
     \setbox0=\vbox{
     \noindent
     \raggedright
     \pretolerance=10000
     {\bf\arabic{Tl}}.\arabic{Tm}.\arabic{Tn}.\
     {\it\ignorespaces#1}\vskip6pt}
     \dimen0=\ht0\advance\dimen0 by\dp0\advance\dimen0 by 4\baselineskip
     \advance\dimen0 by\pagetotal
     \ifdim\dimen0>\pagegoal\eject\fi\fi
     \bgroup
     \raggedright \noindent
     \pretolerance=10000
     {\bf\arabic{Tl}}.\arabic{Tm}.\arabic{Tn}.\
     {\it\ignorespaces#1}\vskip6pt
     \egroup
     \nobreak
     \let\lasttitle=C%
     \parindent=0pt
     \everypar={\parindent=1.5em
     \let\lasttitle=N\everypar={\let\lasttitle=N}}%
     \ignorespaces}
 \def\titled#1{\stepc{To}
     \resetcount{Tp}
     \if N\lasttitle\else\vskip-3pt\vskip-\baselineskip
     \fi
     \vskip 11pt
     \bgroup
     \it
     \noindent
     \ignorespaces#1\unskip. \egroup
     \let\lasttitle=N\ignorespaces}
\let\REFEREE=N
\newbox\refereebox
\setbox\refereebox=\vbox
to0pt{\vskip0.5cm\fullline{\hrulefill\tentt\lower0.5ex
\hbox{\kern5pt referee's copy\kern5pt}\hrulefill}\vss}%
\def\refereelayout{\let\REFEREE=M\footline={\copy\refereebox}%
\message{|A referee's copy will be produced}\par
\if N\lr\else
\if R\lr
\shipout\vbox{\makeheadline
\line{\box\leftcolumn}\makefootline}\advancepageno
\fi\let\lr=N
\topskip=10pt
\output={\plainoutput}%
\fi
}

\newcount\sterne \sterne=0
\newdimen\fullhead
\newtoks\RECDATE
\newtoks\ACCDATE
\newtoks\MAINTITLE
\newtoks\SUBTITLE
\newtoks\AUTHOR
\newtoks\INSTITUTE
\newtoks\SUMMARY
\newtoks\KEYWORDS
\newtoks\THESAURUS
\newtoks\SENDOFF
\newlinechar=`\|
\catcode`\@=\active
\let\INS=N%
\def@#1{\if N\INS $^{#1}$\else\if
E\INS\hangindent0.5\parindent\hangafter=1%
\noindent\hbox to0.5\parindent{$^{#1}$\hfil}\let\INS=Y\ignorespaces
\else\par\hangindent0.5\parindent\hangafter=1
\noindent\hbox to0.5\parindent{$^{#1}$\hfil}\ignorespaces\fi\fi}%
\def\mehrsterne{\advance\sterne by1\global\sterne=\sterne}%
\def\FOOTNOTE#1{\mehrsterne\ifcase\sterne
\or\bfootax \ignorespaces #1\efootax
\or\bfootay \ignorespaces #1\efootay
\or\bfootaz \ignorespaces #1\efootaz\else\fi}%
\def\PRESADD#1{\mehrsterne\ifcase\sterne
\or\bfootax Present address: #1\efootax
\or\bfootay Present address: #1\efootay
\or\bfootaz Present address: #1\efootaz\else\fi}%
\def\maketitle{\paglay%
\def\missing{ ????? }
%
\setbox0=\vbox{\parskip=0pt\hsize=\fullhsize\null\vskip2truecm
\edef\test{\the\MAINTITLE}%
\ifx\test\missing\MAINTITLE={MAINTITLE should be given}\fi
\aTa\ignorespaces\the\MAINTITLE\eTa
\edef\test{\the\SUBTITLE}%
\ifx\test\missing\else\aTb\ignorespaces\the\SUBTITLE\eTb\fi
\edef\test{\the\AUTHOR}%
\ifx\test\missing
\AUTHOR={Name(s) and initial(s) of author(s) should be given}\fi
{\tenpoint\leftskip=40pt\ignorespaces\noindent\the\AUTHOR\vskip5.7pt}
\let\INS=E%
\edef\test{\the\INSTITUTE}%
\ifx\test\missing
\INSTITUTE={Address(es) of author(s) should be given.}\fi
{\ninepoint\leftskip=40pt\ignorespaces\noindent\the\INSTITUTE\vskip11.4pt}%
\edef\test{\the\RECDATE}%
\ifx\test\missing
\RECDATE={$[$the date should be inserted later$]$}\fi
\edef\test{\the\ACCDATE}%
\ifx\test\missing
\ACCDATE={\begpet $[$the date should be inserted later$]$}\fi
{\leftskip=40pt\ninepoint\sl\noindent Received
\ignorespaces\the\RECDATE\unskip; accepted \ignorespaces
\the\ACCDATE\vskip20pt}%
\edef\test{\the\SUMMARY}%
\ifx\test\missing
\SUMMARY={Not yet given}\fi
{\ninepoint\leftskip=40pt\noindent{\bf Abstract{\kern.1em}.}\ \ ---\ \
\ignorespaces\the\SUMMARY\vskip11.4pt}
\edef\test{\the\KEYWORDS}%
\ifx\test\missing
\KEYWORDS={Key words should be inserted}\fi
{\ninepoint\leftskip=40pt\noindent{\bf Key words: }
\ignorespaces\the\KEYWORDS\vskip30pt \ \ }} 
\global\fullhead=\ht0\global\advance\fullhead by\dp0
\global\advance\fullhead by12pt\global\sterne=0
{\parskip=0pt\hsize=19.5cc\null\vskip2truecm
\edef\test{\the\SENDOFF}%
\ifx\test\missing\else\insert\footins{\smallskip\noindent\ninepoint
{\it Send offprint requests to\/}: \ignorespaces\the\SENDOFF}\fi
\hsize=\fullhsize
\edef\test{\the\MAINTITLE}%
\ifx\test\missing\message{|Your MAINTITLE is missing.}%
\MAINTITLE={MAINTITLE should be given}\fi
\aTa\ignorespaces\the\MAINTITLE\eTa
\edef\test{\the\SUBTITLE}%
\ifx\test\missing\message{|The SUBTITLE is optional.}%
\else\aTb\ignorespaces\the\SUBTITLE\eTb\fi
\edef\test{\the\AUTHOR}%
\ifx\test\missing\message{|Name(s) and initial(s) of author(s) missing.}%
\AUTHOR={Name(s) and initial(s) of author(s) should be given}\fi
{\tenpoint\leftskip=40pt\ignorespaces\noindent\the\AUTHOR\vskip5.7pt}
\let\INS=E%
\edef\test{\the\INSTITUTE}%
\ifx\test\missing\message{|Address(es) of author(s) missing.}%
\INSTITUTE={Address(es) of author(s) should be given.}\fi
{\ninepoint\leftskip=40pt\ignorespaces\noindent\the\INSTITUTE\vskip11.4pt}%
\edef\test{\the\RECDATE}%
\ifx\test\missing\message{|The date of receipt should be inserted
later.}%
\RECDATE={$[$the date should be inserted later$]$}\fi
\edef\test{\the\ACCDATE}%
\ifx\test\missing\message{|The date of acceptance should be inserted
later.}%
\ACCDATE={$[$the date should be inserted later$]$}\fi
{\leftskip=40pt\ninepoint\sl
\noindent Received \ignorespaces\the\RECDATE\unskip; accepted \ignorespaces
\the\ACCDATE\vskip20pt}%
\edef\test{\the\SUMMARY}%
\ifx\test\missing\message{|There is no Summary.}%
\SUMMARY={Not yet given.}\fi
{\ninepoint\leftskip=40pt\noindent{\bf Abstract{\kern.1em}.}\ \ ---\ \
\ignorespaces\the\SUMMARY\vskip11.4pt}
\edef\test{\the\KEYWORDS}%
\ifx\test\missing\message{|Missing keywords.}%
\KEYWORDS={Not yet given.}\fi
{\ninepoint\leftskip=40pt\noindent{\bf Key words: }
\ignorespaces\the\KEYWORDS\vskip30pt}} 

\edef\test{\the\THESAURUS}%
\ifx\test\missing\THESAURUS={missing; you have not inserted them}%
\message{|Thesaurus codes are not given.}\fi
\if M\REFEREE\let\REFEREE=Y
\normalbaselineskip=2\normalbaselineskip
\normallineskip=2\normallineskip\normalbaselines\fi
\global\sterne=0
\catcode`\@=12
\tenpoint
\let\lasttitle=A}

\def\ut#1{\mathop{\vtop{\ialign{##\crcr
     $\hfil\displaystyle{#1}\hfil$\crcr\noalign
     {\kern1pt\nointerlineskip}\hbox{$\hfil\sim\hfil$}\crcr
     \noalign{\kern1pt}}}}}
\def\undersim{\ut}
\def\lappreq{\undersim{<}}

\MAINTITLE={ Integrated photometric properties of open clusters}
%
\AUTHOR={ P. Battinelli@1, A. Brandimarti@2, and R. Capuzzo--Dolcetta@2 }
\INSTITUTE={@1 Osservatorio Astronomico di Roma, viale del Parco Mellini 84,
I-00136, Roma, Italy; e--mail: 40061::battinelli
             @2 Istituto Astronomico Univ. di Roma, \lq La Sapienza\rq~,
via G.M. Lancisi 29, I-00161, Roma, Italy; e--mail: 40058::dolcetta }

\RECDATE={20-4-1993}
\ACCDATE={24-8-1993}
\SUMMARY={
Galactic open clusters provide an abundant sample of stellar aggregates
of various sizes, ages and metal abundances, apt to constitute a template for
comparison with star systems in other galaxies.
In this paper we present and discuss a standard methodology to synthesize
{\sl U,B,V}
fluxes and colours, and apply it to a set of 138 open clusters. Results
are compared with previous available ones in the literature.
We were able to calibrate a mass--luminosity relation by which
we evaluated the mass of $\simeq 400$ open clusters, leading
to a well defined present--day mass function. The number--complete
sample of galactic open clusters presented in Battinelli \& Capuzzo--Dolcetta
(1991) is enlarged of a $15\%$.}
\KEYWORDS={ Clusters:open and associations}
\SENDOFF={P.Battinelli}
%
\maketitle
\titlea {Introduction}

It is well known that in our Galaxy the two families of open and
globular clusters are quite well distinct in sizes, spatial
distribution, metallicity content, kinematical properties and
even ages. In the galaxies of the Local Group, such a sharp
distinction is no longer possible: as an example, the
popoulous family of stellar clusters in the Magellanic Clouds
spans with continuity a huge interval of masses, ages,
heavy elements content, without any evident distinction in classes.
Therefore, any meaningful comparison between stellar systems
in galaxies requires a careful definition of the sample
characteristics.
\newline Moreover, the data relative to crowded star clusters in galaxies
are mainly relying on their integrated light. As a consequence,
the possibility of a valid comparison between stellar populations
in star clusters of different galaxies  stands also
on an uniform and objective way to compute integrated fluxes
and colours for a reliable sample of galactic open clusters.
In this regard, we note that, in the literature,  various sets of integrated
fluxes and colours for galactic open clusters have been presented (see, e.g.,
Gray 1965; Sagar, Joshi \& Sinvhal 1983, hereafter SJS;
Spassova \& Baev 1986, hereafter SB; Lyng\aa~ 1987; Pandey et al. 1989).
Of course, it is important to know precisely the way followed to compute
the integrated fluxes; unfortunately the richest sample of data
(Lyng\aa~ 1987) is not accompanied by any description of the method.
In particular, it is quite clear the crucial importance of
the selection of bright stars as member of a cluster when
determining its integrated magnitude.
\newline Once the relevant integral quantities are computed,
another crucial point (already addressed in Battinelli \&
Capuzzo--Dolcetta 1991, hereafter BCD) is the
statistical completeness of the sample of clusters to use
as a comparison template.
\par The main aim of this paper is to present and discuss a {\it standard}
method to reliably evaluate integrated magnitudes and colours
(and some derived quantities) for galactic open clusters;
the FORTRAN package  able to produce synthetic
{\sl U,B,V}, fluxes following the method described in Sect. 2.1.
This package can be easily adapted to deal with input data coming from
different sources; this is of particular importance due to the rapidly
increasing
availability of digitized data in astronomical data centers (see for instance
the open cluster data by Mermilliod 1986, available at the Strasbourg data
Centre)

\titlea{Integrated fluxes}
\titleb{ T{\sc HE STANDARD METHOD OF COMPUTATION}}
Our methodology to evaluate {\sl U,B,V} integrated fluxes for galactic open
clusters is  simple: using the most suitable and up--to--date data
(HR diagrams, radial velocities and proper motions when available)
for open clusters available in
the literature, we try to define a standard procedure for
membership determination and for flux summation.
Such a procedure (described below) has been automatized, after the necessary
tests.
\par
Standard assumptions made throughout this work are: the reddening
is the same for all the stars of a given cluster, and was taken from Lyng\aa~
1987; the ratio
$A_V/E_{B-V}$ is set to 3.1; when $E_{U-B}$ is not available,
the relation
$E_{U-B}=0.72E_{B-V}$ is assumed; when the data available
refer to the $RGU$ system (Becker 1946) they are transformed to $UBV$
via the relations given by Steinlin (1968) (this happened for only
10 clusters).
\par
Regarding to membership, we relied on quoted references and decided to
use criteria:
i) to exclude stars just when explicitly suggested by radial velocity
or proper motion in the source paper;
ii) to exclude stars external to a certain angular field,
as given in the reference quoted, which is usually defined
on the basis of stellar counts;
iii) to exclude very bright blue stars just when red
giant are surely members, and just when they are at least
$3~mag$ brighter than the turn--off (this to avoid to exclude
possible blue--stragglers, which have been found in many
clusters). Anyway, it cannot be ruled out the possibility
that a population of blue bright stars is a real feature of the cluster,
even when the presence of evolved red giants seems in
contradiction to that. Indeed, for example, bright blue stars can be part of
a second generation. For this reason, the exclusion of
blue bright stars is done only when they are very few ($\leq 2$).
\par
In addition to the membership determination,
the main sources of error in the evaluation of the
absolute synthetic magnitudes are: the photometry of the single stars
(standard errors of $\approx 0.02$ and $0.06~mag$,
for photoelectric and photographic
data, respectively); the distance modulus determination, whose
typical error is $\lappreq 0.4~mag$ (see SJS);
the reddening, carrying an indeterminacy
of $\sim \pm 0.1~mag$ (see SJS), when
the hypothesis of constant $E_{B-V}$ inside the cluster is
acceptable; the incoming incompleteness of the HR diagram
of the cluster at faint magnitudes, whose importance is weighted by the
slope of the luminosity function.
\newline Globally, we can confirm the SJS conclusion that
the maximum error in magnitude is about $0.5~mag$ and about $0.2~mag$ in
colour.
\titleb{ T{\sc HE COMPARISON OF OUR RESULTS WITH AVAILABLE DATA}}
The Lyng\aa's~ {\it Catalogue of Open Cluster Data}
(1987)
contains values of integrated {\sl V} and $B-V$ for
$\approx 400$ open clusters, mostly computed by Skiff.
Unfortunately this data--base of synthetic fluxes,
 which is the most abundant available,
 has not been accompanied by a description
of the methodology used and  sometimes it is even difficult to identify
the original source of data.
Anyway, the abundance  of Lyng\aa's data suggests to check them with fluxes
and colours obtained with
an independent method whose reliability can be evaluated.
In this regard,
we decided to perform the comparison on a representative
sub--set of the Lyng\aa's
sample consisting of clusters spanning large intervals of age, distance,
and richness.
\begfig 4.25 cm
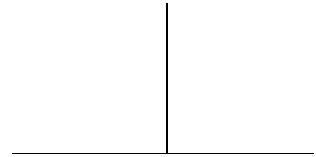
\figure{1} {Comparison between our (abscissa) and Lyng\aa's integrated
$M_V$ (panel a) and $(B-V)_0$ (panel b).
The sample of 53 clusters is composed only by clusters whose
data are taken from the same reference source used by Lyng\aa.
Squares indicate clusters of the BCD {\it complete} sample.}
\endfig
Figure 1 shows the comparison between our and Lyng\aa's
integrated magnitudes and colours for a sample of $N=53$ clusters
adopting the same sources of data.
The agreement is good (the slope of the  $M_V$ correlation
is $a=0.99\pm 0.03$, with a correlation coefficient $r=0.98$;
$a=0.92\pm 0.05$ and  $r=0.93$ for $(B-V)_0$),
particularly for the bright and nearby clusters of the
BCD sample ($M_V\leq -4.5$ and projected heliocentric
distance $d\leq 2~Kpc$). In both our and Lyng\aa's determination some
(anomalous) too blue clusters are present. The reason for that
is both an
excessive average value of $E(B-V)$ as deduced from Lyng\aa's ~catalogue
and the dispersion of individual reddenings around the average. Actually,
we checked that the integral $(B-V)_0$ of Boc 7 (the only cluster of the
7 clusters bluer than $(B-V)_0=-0.5$ for which these data are available)
changes from $-0.6$ to
$-0.24$ when using individual $E(B-V)$. This is due to a
systematically small (respect to the average) reddening
of the luminous stars in Boc 7.
\newline A systematic overestimate of our integral {\sl V}--luminosity with
respect to
Lyng\aa~ is clear at faint magnitudes, and it has been identified
as due to the exclusion by Skiff of some
very bright stars. These exclusions  do not seem to be motivated
by a unique criterion for all clusters.
Obviously, the exclusion of bright
stars is more important for the computation of the integral magnitude
of poor clusters. The difference in the integrated colours
is not systematic, depending on the colour of the excluded
stars.\par
The inclusion of more updated individual cluster data, with
respect to those used by Skiff, obviously increases the scatter
in the integrated fluxes correlation as shown in
Figure 2. \par
\begfig 4.25 cm
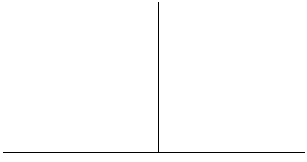
\figure{2} {As in Fig. 1 for clusters of Table 1.}
\endfig
A comparison of our integrated $V$ magnitude, $B-V$ and $U-B$ colours
with those given by Gray (1965), SJS, SB and Pandey et al. (1989) is given in
Figure 3 and 4; the agreement is not as good as in the
case of Lyng\aa's data, due to both different references and  methodology.
\begfig 5.5 cm
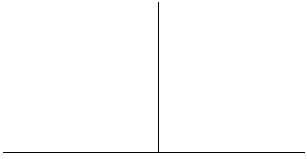
\figure{3}{Comparison between our integrated $M_V$ (abscissa) and those
computed by Gray (1965) (squares), Sagar et al. (1983) and Pandey et al.
(1989) (crosses), and Spassova \& Baev (1985) (triangles).}
\endfig
\begfig 3.5 cm
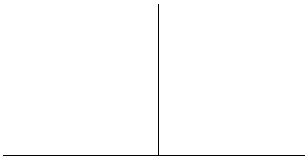
\figure{4}{$(B-V)_0$ and $(U-B)_0$ correlations  (panel a and b) of ours and
Gray (1965), Sagar et al. (1983), Pandey et al. (1989) and Spassova
\& Baev (1985). Symbols as in Figure 3.}
\endfig
Table 1 shows our integrated $U,B,V$ magnitudes and colours for a sample
of clusters composed by all the objects in Lyng\aa's catalogue within the
projected distance (onto the galactic plane)
$d=2~Kpc$ from the Sun ( thus to enlarge the BCD sample) and of a selected set
of clusters apt to check our method versus Lyng\aa's data.
\newline
The HR diagrams of the various clusters, as available from the literature,
are obviously not homogeneous in their completeness characteristics.
For this reason, we have introduced $V_{lim}$ as the faintest magnitude
of stars included in our flux summation.
 Of course, the
importance of $V_{lim}$ is determined by the luminosity
function of the cluster, in the sense that a variation of $V_{lim}$
is more relevant for clusters poor in high luminosity stars and/or
having a steep slope of the luminosity function in the low main sequence.
An estimate good within $\leq 0.05~mag$ in the integral V--magnitude
(corresponding to $\Delta L_V/L_V \simeq 5\%$) requires
to include in the computation of the flux, stars in an apparent magnitude
interval  at least $\Delta V_{min}=6.0$ starting from the brightest member.
\titleb{ T{\sc HE NEW COMPLETE SAMPLE}}
The data base of all clusters with known distance and integrated $M_V$ allows
us to update the number--complete sample described in BCD.  If
we make the hypothesis of uniform averaged distribution of clusters
(i.e. the assumption of a constant number density of clusters within
cylinders of radius $d$ centered to the Sun, see BCD Sect. 2)
\begfig 4.5 cm
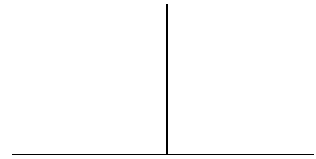
\figure{5}{Count of clusters brighter than $M_V=-4.5$ inside
cylinders of radius $d$ (in $Kpc$) centered at the Sun. The
dashed straight line has slope $2$, to correspond to a uniform
distribution, and is normalized to the observed number of clusters
within $d=1~Kpc$.}
\endfig
Figure 5 shows how the sample of $N=115$ clusters brighter than $M_V=-4.5$ and
with
projected distance  $d\leq 2~Kpc$ is number--complete.
 Hereafter we will
refer to it as  the \lq complete\rq ~sample, whose relevant characteristics
are given in Table 2.
The present--day cluster formation rate as deduced by the counts of clusters
younger than $Log~t(yr)=6.5$ in the new sample is $R=48\times
10^{-8}Kpc^{-2}yr^{-1}$,
i.e. unchanged (within the $10\%$ error estimated in BCD) with respect
to the BCD evaluation.
\titleb{ T{\sc HE HYADES CLUSTER}}
The Hyades are known to constitute a reference cluster apt to give
a reliable fiducial unreddened main sequence and two--colour diagram.
\newline
By mean of the HR diagram (187 members) given by  Hagen (1970)
we get
$V=0.49$, $M_V=-2.91$, $B-V=(B-V)_0=0.40$;
these data are almost coincident with Lyng\aa's values
($V=0.50$, $M_V=-2.91$, $(B-V)_0=0.40$) in spite of the different number
(380) of member stars.
This means   that $\approx 50\%$ of stars in Lyng\aa's
sample contribute negligibly to the total light.
\newline Adopting more recent data (Pels, Oort \& Pels--Kluyver 1975;
Griffin et al. 1988; Gunn et al. 1988) we collected 256 stars as certain
members (on the basis of radial velocities), and  obtained (with
$(M-m)_V=3.24~ mag$ from Gunn et al. 1988) $V=0.56$,
$M_V=-2.68$, $(B-V)_0=0.40$.
\titlea{ Photometric results}

The results of Sect. 2.2  indicate the good degree of reliability of
Lyng\aa's integrated magnitudes.
For this reason, the global results obtained from
his $B,V$ integrated photometry stand almost unaltered by our enlargement
of his sample of $\sim 10\%$. Anyway, our computation of $U$ fluxes for the
in Table 1 suggested us to examine the photometric characteristics of our
sample.
\par Figure 6  is the integrated colour--magnitude diagram of
the clusters of Lyng\aa's catalogue and of Table 1 sample.
This figure seems to confirm the reliability of cluster integrated photometry
because just  few clusters lie  below
the luminosity class V (taken from Allen 1973).
\begfig 4.5 cm
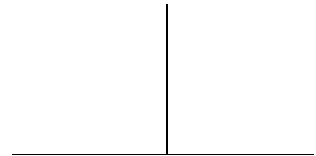
\figure{6}{Colour--magnitude diagram for the whole set
of ours and Lyng\aa's clusters. The lower and upper solid lines are theoretical
evolutionary tracks (see text) of a solar composition cluster of total mass
$25~M_\odot$  and $40,000~M_\odot$, respectively. The thick curve
is the location of the luminosity class V.}
\endfig
The colour--colour diagram in Figure 7
shows a defined trend with a slope in good agreement with the $Z=0.02$
theoretical model taken from Battinelli and Capuzzo--Dolcetta (1989).
\begfig 4.5 cm
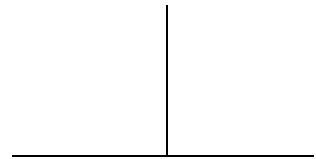
\figure{7}{This colour--colour diagram refers to clusters of Table 1.
The thin curve is the theoretical model, while the thick one
refers to luminosity class V.}
\endfig
\newline Note that in Figures 6 and 7 some clusters are present with
anomalously blue ($(B-V)_0\leq 0.5$) colour. Five clusters out of six are
poor and distant; the reason for their too blue colours is both an
excessive average value of $E(B-V)$ as deduced from Lyng\aa's ~catalogue
(or, equivalently, the photometry has a blue bias)
and the dispersion of individual reddenings around the average. Actually,
we checked that the integral $(B-V)_0$ of Boc 7 (the only cluster of the
6 for which these data are available) changes from $-0.6$ to
$-0.24$ when using individual $E(B-V)$. This is due to a
systematically small (respect to the average) reddening
of the luminous stars in Boc 7.
\par
Figure 8 shows the $M_V$ vs age relation; let us note that clusters lie between
the theoretical curves corresponding to a 25
$M_\odot$ and $4\times10^4$
$M_\odot$ cluster, as obtained using the theoretical model above.
\begfig 9 cm
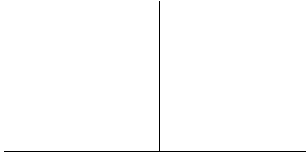
\figure{8}{Panel a: integrated $M_V$ vs age relation for the same sample of
Fig. 6.
Panel b: integrated $(U-B)_0$ vs age for the same sample of Fig. 7.
The curves are the same theoretical models of Fig. 6.}
\endfig
As it happens for Magellanic Cloud clusters (see Battinelli \&
Capuzzo--Dolcetta 1989) we find that
$(U-B)_0$ is slightly better correlated to the cluster
age than $(B-V)_0$, the least--square relation being:
$$(U-B)_0=0.468 Log~t-3.95,\eqno\autnum$$
the correlation coefficient is $r=0.643$.
\titlea{ A photometric estimate of open cluster masses}

The best way to evaluate the total mass of a stellar system is,
in principle, given by dynamical considerations based on velocity
and position data of member stars.
However, suitable radial velocity and proper motion data are available
just for 9 clusters, and so it may be helpful to resort to photometric data.
A photometric estimate of the total mass of a cluster is (see also
Pandey, Bhatt \& Mahra 1987, hereafter PBM)
$$\eqalign{M=& \int_0^\infty m\psi (m)dm \simeq \int_{m_{lim}}^\infty m\psi
(m)dm
= \cr
& = \int^{\infty}_{L_{Vlim}} m(L_V)\phi (L_V)dL_V} \eqno\autnum$$
(where $\psi(m)$ and $\phi(L_V)$ are, respectively, the mass and
$V$--luminosity functions of the cluster, and $m_{lim}$ is the mass
corresponding to $L_{Vlim}$).
Of course, this evaluation of cluster mass stands on a suitable
mass--luminosity
relation; moreover, $L_{Vlim}$ should be low enough to allow
the convergence of the last integral in Eq. (2) (hereafter mass and
luminosities
are in solar units).
\par The approximations adopted are: i) to main--sequence stars we assigned
their ZAMS mass; ii) to post--main sequence stars we assigned the turn--off
mass value. The relative error introduced is evaluated to be less than
$10\%$.
The mass--luminosity relation on the ZAMS has been calibrated
on the basis of Mengel et al. 1979 ($0.7\leq m/M_\odot\leq 3$)
and Becker 1981 ($3\leq m/M_\odot\leq 9$) sets of data suitably matched
in the region overlap.
A solar composition
($X=0.70, Y=0.28, Z=0.02$) was chosen; this required interpolations
in $Y$ between 0.2 and 0.3 and in $Z$ between 0.01 and 0.04 of
Mengel et al. data. To give a rough evaluation of the sensitivity
of the mass on the ZAMS at varying $Y$ and $Z$ in the mentioned
intervals we found $\left| \Delta m/m\right|\simeq \left| \Delta M/M\right|
\lappreq
10\%$ when $Y$ varies and $Z$ is kept constant, and
$\left| \Delta m/m\right|\simeq \left| \Delta M/M\right| \lappreq
20\%$ when $Z$ varies and $Y$ is kept constant.
In some cases it was needed to extrapolate
the $m-L$ relation outside the $0.7-9 M_\odot$ interval.
A high mass extrapolation was required for 4 clusters only;
however, because of the scarcity of these massive ($m>9 M_\odot$) stars,
the evaluation of the total cluster mass is not sensibly
affected. An extrapolation below $0.7~M_\odot$ was necessary
just for Hyades, where independent estimates of the mass allow
to check the error.
\newline The bolometric corrections necessary to convert visual
into bolometric luminosities have been taken
from Johnson (1966) and Morton and Adams
(1968). Note that the bolometric correction induces an error small
in comparison with that due to other parameters (e.g. the distance modulus).
With regard to the influence of a variation of the low--luminosity
cut--off, $L_{Vlim}$, we checked that to have a relative error less than
$5\%$ in the cluster mass estimate, all the stars in the range of at
least 9 $mag$ fainter than the brightest cluster star should be included.
\begfig 7 cm
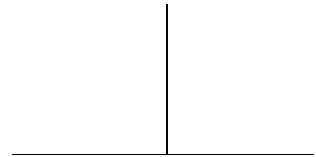
\figure{9}{Correlation of our mass evaluations (abscissa) and those by
Schmidt (1963) (small black squares), Pandey et al. (1987) (black
squares), and Bruch \& Sanders (1983) (empty squares). Masses are in solar
units.}
\endfig
\par Figure 9 shows a comparison between our mass evaluations and those
by Schmidt (1963), PBM, and
Bruch \& Sanders (1983). Our values are systematically lower than Schmidt's
ones (which, being dynamical estimates, take also into account non--luminous
mass) and, in the average, slightly larger than PBM ones. Probably, PBM values
are smaller than ours due
to incoming incompleteness at their fainter magnitudes causing
an apparent convergence of the total mass as given by Eq. (2).
\newline
The slope of the least square fit to the
relation between Schmidt's and our masses is close to 1:
this indicates (even with a large dispersion) the remarkable result of
a roughly constant fraction of
non--luminous mass to the total in open clusters.
\par For the sake of comparison, Figure 10 shows our and PBM
cumulative masses in function of the individual mass for the Hyades cluster,
for which we used the detailed set of data by Griffin et al. (1988).
PBM give $78 M_\odot$ for the total luminous Hyades mass;
we obtain $256\pm 3 M_\odot$, against the $692M_\odot$ given by Schmidt and
$390M_\odot$ by Gunn et al. 1988, who both consider non--luminous
matter.
\begfig 4.5 cm
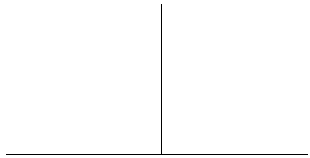
\figure{10}{Value of the mass computed via Eq.(2) in function of $m_{lim}$
for the Hyades cluster. Crosses: this paper; dots: Pandey et al. (1987).
Masses are in solar
units.}
\endfig
Figure 10 is an example of how the existence of a plateau in the
cumulative mass frequency may be due to incompleteness, rather than to
the actual convergence of the integral in Eq (2).
\newline
The mass--luminosity relation for clusters in our sample is shown
in Figure 11; the least--square fit to this relation is:
$$ LogM=-0.15M_V+1.43, \eqno\autnum$$
with correlation coefficient $r=0.69$. The standard deviation of data
from the estimate is $\sigma=0.29$, corresponding to a factor
$\approx 2$ in the mass.
\begfig 5.5 cm
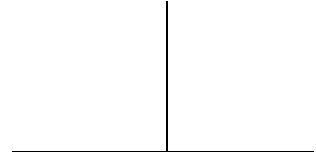
\figure{11}{Logarithmic mass--luminosity relation for clusters of Table 1. The
least square fitting relation is shown. Masses are in solar
units.}
\endfig
An increasing trend of
the $M/L_V$ ratio with age is found even with a large dispersion
(see Figure 12); the corresponding least square fit
$$ Log{M\over L_V}=0.439Logt-5.09, \eqno\autnum$$
with r=0.67.
\begfig 4.25 cm
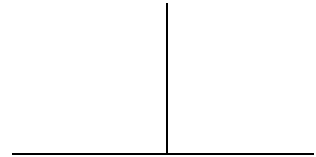
\figure{12}{$M/L_V$ ratio vs age (logarithmic scales) for our computed masses
in Table 1.}
\endfig
Individual values of cluster mass deduced via Eq. (4) are suitable
just at the level allowed by the large dispersion in the data; anyway,
it can be reliably used to obtain the statistical information
given by the global mass function.
The resulting mass histogram of all clusters in Lyng\aa's catalogue with
available age and luminosity is shown
in Figure 13.
\begfig 6.25 cm
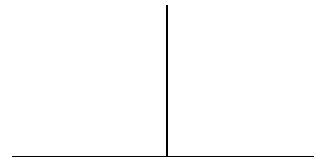
\figure{13}{Histograms of the cluster masses for the whole set of clusters
(see Fig. 6) and of the sample of Table 2.}
\endfig
The exponent of the least--square fitting to the decreasing part
of the mass function is $-2.04\pm 0.11$ ($r=0.96$) and $-2.13\pm 0.15$
($r=0.96$)
for our complete sample. These values compare with $-2.2$ given by Reddish
(1978), who included associations, too.
The peak value of our mass function is at $126 M_\odot$, instead of
$\sim 360~M_\odot$ of Reddish (1978).
\titlea{Conclusions}

In this paper we presented a {\it standard} method to synthesize
in a reliable and automatic way the
{\sl U,B,V} luminosities of stars in an  open cluster. One of the most
important parts of the work is the
selection of member stars based, as usual, on
radial velocities and proper motions (when available) and HR diagrams.
We show how a convergence within $0.05$ magnitudes of the
integral magnitude is reached when all the stars in the HR diagram
down to $6~mag$ fainter than the brightest member are included.
\par
The comparison of our photometric results for a sample of clusters for
which integral $B$ and $V$ data are found in Lyng\aa's 5th Catalogue of
Open Cluster Data indicates a high degree of compatibility between the
two sets of data. This makes us confident about the
use of the synthetic fluxes given by Lyng\aa's catalogue (whose
integration procedure is not discussed) and, consequently, about our
enlargement of that sample. An important new contribution is the
evaluation of $U$ magnitudes and $U-B$ colours for 115 clusters.
\par We also discussed a method to evaluate the luminous mass
of a cluster via the computation of its cumulative mass distribution;
this allowed us to calibrate a mass--luminosity relation which
was used to obtain the present-day mass function of all the
clusters with available integral $M_V$ and age.
The exponent of the high mass tail of the mass function is
$-2.04$, and it represents an updating (being also based on a much more
abundant sample) of Reddish's (1978)
$-2.2$ result. This mass function is flatter than that of stars
in the solar neighbourhood. A comparison of our mass estimates (referring
to luminous stars in a cluster) with the dynamical estimates by Schmidt (1963)
yields to a roughly constant ratio of non--luminous to the total
mass in the cluster sample examined.
Finally, the new integral magnitudes
presented here lead to an increase up to 115 members of the
set of clusters brighter than $M_V=-4.5$ and inside the $2~Kpc$
radius cylinder around the Sun, i.e. the number--complete sample
defined in Battinelli \& Capuzzo--Dolcetta (1991). This sample
constitutes a natural luminosity--unbiased template for a comparison
with star clusters in galaxies and gives precise constraints to
evolutive models of the open cluster system in our galaxy.

%
%
%
\begref

\ref Allen, C.W., 1973, Astrophysical Quantities, 3rd ed.,
(London: Athlone Press)
\ref Battinelli, P., Capuzzo--Dolcetta, R., 1989, ApJ 347, 794
\ref Battinelli, P., Capuzzo--Dolcetta, 1991, R., MNRAS 249, 76
\ref Becker, A.S., 1981, ApJS 45, 475
\ref Becker, W., 1946, Ver\"off. Univ. Sternwarte G\"ottingen, n.79
\ref Bruch, A. \& Sanders, W.L., 1983, A\&A 121, 237
\ref Gray, D.F, 1965, AJ 70, 362
\ref Griffin, R. F., Gunn, J. E., Zimmermann, B. A. \& Griffin, \hfill
R.E.M., 1988 AJ, 96, 172
\ref Gunn, J. E., Griffin, R. F., Griffin, R.E.M. \& Zimmermann, B.A., 1988,
AJ 96, 198
\ref Johnson, H.L., 1966, ARA\& A 4, 193
\ref Hagen, G.L., 1970, Publ. David Dunlap Obs., Univ. Toronto 4,1
\ref Lyng\aa~ , G., 1987, Lund Catalogue of Open Cluster Data, 5th ed.,
Stellar Data Centre, Observatoire de Strasbourg, France
\ref Mengel, J.G, Sweigart, A.W., Demarque, P. \& Gross, P.G., 1979,
ApJS 40, 733
\ref Mermilliod, J.C., 1986, AAS 24, 159
\ref Morton, D.C. \& Adams, T.F., 1978, ApJ 151, 611
\ref Pandey, A.K., Bhatt, B.C. \& Mahra, H.S., 1987, ApSS 129, 293
\ref Pandey, A.K., Bhatt, B.C., Mahra, H.S. \& Sagar, R., 1989,
MNRAS 236, 263
\ref Pels, G., Oort, J.H. \& Pels--Kluyver, H.A., 1975, A\&A 43,423
\ref Reddish, V.C., 1978, Stellar Formation, (Oxford: Publ. Pergamon Press)
\ref Sagar, R., Joshi, U.C. \& Sinvhal, S.D., 1983, Bull. Astron. Soc.
India 11, 44
\ref Schmidt, K.H., 1963, Astron. Nach. 287, 41
\ref Spassova, N.M. \& Baev, P.V., 1985, A\&AS 112, 111
\ref Steinlin, U.W., 1968, Z. Astrophys. 69, 276
\endref
\vfill\eject\noindent
\titlea{Appendix}

References to Table 1:
\par\noindent
\reff (1) Alcala, J. M. \& Ferro, A. A., 1988, Rev. Mex. Astron. Ap., 16, 2,
81\reff
(2) Battinelli, P., Capuzzo--Dolcetta, R. \& Nesci, R., 1992, AJ, 103, 5,
1596\reff
(3) Becker, W., 1960, Zeit. fur Astrophysik, 51, 49\reff
(4) Bohm-Vitense, E. et al., 1984, ApJ, 287, 2, 825\reff
(5) Claria, J. J., 1972, A\&A, 19, 303\reff
(6) Claria, J. J., 1977, PASP, 89, 803\reff
(7) Claria, J. J. \& Lapasset, E., 1985, MNRAS, 214, 2, 229\reff
(8) Claria, J. J. \& Lapasset, E., 1988, MNRAS, 235, 4, 1129\reff
(9) Claria, J. J. \& Lapasset, E., 1989, MNRAS, 241, 2, 301\reff
(10) Cudworth, K.M., 1976, A\&AS, 24, 143
\reff (11) Delgado, A. J. et al., 1984, A\&AS, 58, 3, 447
\reff (12) Delgado, A. J. et al., 1992, AJ, 103, 3, 891
\reff (13) Dupuy, D L \& Zukauskas, W., 1976, Roy. Astron. Soc. Canadà J., 70,
169
\reff (14) Eggen, O. J., 1972, AJ, 173, 63
\reff (15) Eggen, O. J., 1981, ApJ, 247, 507
\reff (16) Evans, T. L., 1969, MNRAS, 146, 101
\reff (17) Fehrenbach, Ch. et al., 1987, A\&AS, 68, 515
\reff (18) Feinstein, A., 1964, The Observatory, 940, 111
\reff (19) Feinstein, A. et al., 1973, A\&AS, 9, 233
\reff (20) Feinstein, A. et al., 1978, A\&AS, 34, 241
\reff (21) Feldbrugge \& van Genderen, 1973, A\&AS, 91, 209
\reff (22) Fenkart, RP \& Schroder, A., 1985, A\&AS, 59, 1, 83
\reff (23) Fernandez, J. A. \& Salgado, W., 1980, A\&AS, 39, 11
\reff (24) FitzGerald, M. P. et al., 1978, A\&AS, 37, 351
\reff (25) FitzGerald, M. P. et al., 1990, PASP, 102, 865
\reff (26) Forbes, D. et al., 1992, AJ, 103, 3, 916
\reff (27) Gatewood, G. et al., 1988, ApJ, 332, 2, 917
\reff (28) Gieseking, F., 1981, A\&A, 99, 155
\reff (29) Gieseking, F., 1985, A\&AS, 61, 75
\reff (30) Girard, T. M. et al., 1989, AJ, 98, 1, 227
\reff (31) Glaspey, J. W., 1987, PASP, 99, 1089
\reff (32) Griffin, R. F. et al., 1988, AJ, 96, 1, 172
\reff (33) Grubissich, C., 1973, A\&AS, 11, 287
\reff (34) Gunn, J. E. et al., 1988, AJ, 96, 1, 198
\reff (35) Hagen, G. L., 1970, Publ. David Dunlap Obs. Univ. Toronto Vol 4
\reff (36) Hassan, S. M.,1975, A\&AS, 20, 255
\reff (37) Hill, G. \& Perry, C. L., 1969, AJ, 74, 7, 1011
\reff (38) Hiltner, W. A. et al., 1965, ApJ, 141, 1, 183
\reff (39) Hoag, A. A. et al., 1961, Publ. U. S. Naval Obs. 2 ed., 17, 343
\reff (40) Ianna, P. A. et al., 1987, AJ, 92, 2, 347
\reff (41) Johansson, K. L. V., 1981, A\&AS, 43, 421
\reff (42) Johnson, H. L. \& Morgan, W. W., 1953, ApJ, 117, 3, 313
\reff (43) Jones, B. F. \& Cudworth, K., 1983, AJ, 88, 2, 215
\reff (44) Joshi, U. C. \& Sagar, R., 1976, ApSS, 48, 225
\reff (45) Joshi, U. C. \& Sagar, R., 1983, MNRAS, 202, 961
\reff (46) Koelbloed, D., 1959, BAN, 14, 489, 265
\reff (47) Landolt, A. U. et al., 1990, AJ, 100, 8, 695
\reff (48) Lee, P. D. \& Perry, C. L., 1972, AJ, 76, 5, 464
\reff (49) Levato, H. et al., 1988, ApSS, 146, 2, 361
\reff (50) Lindoff, U., 1971, A\&A, 15, 439
\reff (51) Loden, L. O., 1983, A\&AS, 53, 1, 33
\reff (52) Loden, L. O., 1984, A\&AS, 58, 3, 595
\reff (53) Lyng\aa,~ G., 1990, A\&A, 54, 311
\reff (54) Marschall, L. A. et al., 1990, AJ, 99, 5, 1536
\reff (55) Mathieu, R. D. et al., 1986, AJ, 92, 5, 1100
\reff (56) Mathys, G., 1991, A\&A, 245, 467
\reff (57) Mc Cuskey,S. W. \& Houk, N., 1964, AJ, 69, 6, 412
\reff (58) McNamara, B. J. et al., 1989, AJ, 97, 5, 1427
\reff (59) Mermilliod, J. C., 1976, A\&AS, 24, 159
\reff (60) Mermilliod, J. C. et al., 1990, A\&A, 235, 114
\reff (61) Moffat, A. F. J., 1974, A\&AS, 16, 33
\reff (62) Moffat, A. F. J. \& Vogt, N., 1973, A\&AS, 10, 135
\reff (63) Moffat, A. F. J. \& Vogt, N., 1973, A\&AS, 11, 3
\reff (64) Moffat, A. F. J. \& Vogt, N., 1975, A\&AS, 20, 85
\reff (65) Moffat, A. F. J. \& Vogt, N., 1975, A\&AS, 20, 125
\reff (66) Moffat, A. F. J. \& Vogt, N., 1975, A\&AS, 20, 155
\reff (67) Morgan, W. W. \& Hiltner, W. A., 1965, ApJ, 141, 1, 177
\reff (68) Pedreros, M., 1987, AJ, 94, 5, 1237
\reff (69) Pels, G. et al., 1975, A\&A, 43, 423
\reff (70) Perez, M. R. et al., 1987, PASP, 99, 1050
\reff (71) Perry, C. L. \& Hill, G., 1969, AJ, 74, 7, 899
\reff (72) Perry, C. L. et al., 1978, PASP, 90, 73
\reff (73) Perry, C. L. et al., 1978, PASP, 90, 81
\reff (74) Perry, C. L. et al., 1990, A\&AS, 86, 415
\reff (75) Perry, C. L. et al., 1991, A\&AS, 90, 2, 195
\reff (76) Peterson, J. C. \& FitzGerald, M. P.,1988, MNRAS, 235, 1439
\reff (77) Prosser, C. F.,1992, AJ, 103, 2, 489
\reff (78) Rahim, M. A., 1966, Astron. Nachr., 289, 75
\reff (79) Richtler, T., 1985, A\&AS, 59, 3, 491
\reff (80) Richtler, T. \& Kaluzny, J., 1989, A\&AS, 81, 2, 225
\reff (81) Sanders, W. L., 1977, A\&AS, 27, 89
\reff (82) Sanders, W. L., 1990, A\&AS, 84, 615
\reff (83) Sanduleak, N. \& Philip, A. G. D., 1968, AJ 73, 7, 566
\reff (84) Seggewiss, W., 1968, Zeit. fur Astrophysik, 68, 142
\reff (85) Stauffer, J. et al., 1989, ApJ, 342, 1, 285
\reff (86) Steinlin, U., 1960, Zeit. fur Astrophysik 50, 233
\reff (87) Stock, J., 1984, Rev. Mex. Astron. Ap., 9, 2, 127
\reff (88) Svolopoulos, S N, 1966, Zeit. fur Astrophysik, 64, 67
\reff (89) Talbert, F. D., 1975, PASP, 87, 341
\reff (90) The, P. S. \& Stokes, N., 1970, A\&A, 5, 298
\reff (91) Turner, D. G., 1986, AJ, 92, 1, 111
\reff (92) Turner, D. G. \& Pedreros, M., 1985, AJ, 90, 7, 1231
\reff (93) Turner, D. G. et al., 1983, AJ, 88, 8, 1199
\reff (94) van Genderen, A. M. et al., 1984, A\&AS, 58, 3, 537
\reff (95) Vasilevskis, S. et al., 1965, AJ, 70, 10, 797
\reff (96) Vazquez, R. A. \& Feinstein, A., 1990, A\&AS, 86, 209
\reff (97) Vazquez, R. A. \& Feinstein, A., 1991, A\&AS, 87, 2, 383
\reff (98) Verschoor, J. N. \& van Genderen, A. M., 1983, A\&AS, 53, 419
\reff (99) Vogt, N. \& Moffat, A. F. J., 1972, A\&AS, 7, 133
\reff (100) Vogt, N. \& Moffat, A. F. J., 1973, A\&AS, 9, 97
\reff (101) Walker, A. R., 1985, MNRAS, 213, 4, 889
\reff (102) Walker, A. R. \& Laney, C. D., 1987, MNRAS, 224, 1, 61
\reff (103) Walker, M. F., 1956, ApJS, 2, 23, 365
\reff (104) Westerlund, B. E. et al., 1988, A\&AS, 76, 1, 101
\reff (105) Wizinowich, P. \& Garrison, R.F., 1982, AJ, 87, 10, 1390
\reff (106) Yilmza, F., 1970, A\&A, 8 213
\bye